\documentclass[prd,superscriptaddress,showpacs,nofootinbib,amsmath,amssymb,aps,11pt]{revtex4}

\usepackage{bm}
\usepackage{amsfonts}
\usepackage{latexsym}
\usepackage[latin1]{inputenc}
\usepackage{graphicx}
\usepackage{amsmath}
\usepackage{palatino}
\usepackage{mathpazo}
\usepackage{booktabs}
\usepackage{dcolumn}

\def\jnl@style{\it}
\def\aaref@jnl#1{{\jnl@style#1}}

\def\aaref@jnl#1{{\jnl@style#1}}

\def\aj{\aaref@jnl{AJ}}                   
\def\apj{\aaref@jnl{ApJ}}                 
\def\apjl{\aaref@jnl{ApJ}}                
\def\apjs{\aaref@jnl{ApJS}}               
\def\apss{\aaref@jnl{Ap\&SS}}             
\def\aap{\aaref@jnl{A\&A}}                
\def\aapr{\aaref@jnl{A\&A~Rev.}}          
\def\aaps{\aaref@jnl{A\&AS}}              
\def\mnras{\aaref@jnl{Mon.~Not.~Roy.~Astron.~Soc.}}             
\def\prd{\aaref@jnl{Phys.~Rev.~D}}        
\def\prc{\aaref@jnl{Phys.~Rev.~C}}  
\def\prl{\aaref@jnl{Phys.~Rev.~Lett.}}    
\def\qjras{\aaref@jnl{QJRAS}}             
\def\skytel{\aaref@jnl{S\&T}}             
\def\ssr{\aaref@jnl{Space~Sci.~Rev.}}     
\def\zap{\aaref@jnl{ZAp}}                 
\def\nat{\aaref@jnl{Nature}}              
\def\aplett{\aaref@jnl{Astrophys.~Lett.}} 
\def\apspr{\aaref@jnl{Astrophys.~Space~Phys.~Res.}} 
\def\physrep{\aaref@jnl{Phys.~Rep.}}      
\def\physscr{\aaref@jnl{Phys.~Scr}}       
\def\commat{\aaref@jnl{Comm.~Math.~Phys.}}              
\def\science{\aaref@jnl{Science}}               
\def\cqg{\aaref@jnl{Classical Quant.~Grav.}}            
\def\jpcs{\aaref@jnl{JPCS}}                                     
\def\ijmpd{\aaref@jnl{Int.~J.~Mod.~Phys.~D}}                    
\def\grg{\aaref@jnl{Gen.~Relat.~Gravit.}}               
\def\rpp{\aaref@jnl{Rep.~Prog.~Phys.}}          
\def\npa{\aaref@jnl{Nucl.~Phys.~A}}        
\def\lrr{\aaref@jnl{Living Rev.~Rel.}}                   
\def\jcap{\aaref@jnl{J.~Cosmology Astropart.~Phys.}}    
\def\rmp{\aaref@jnl{Rev.~Mod.~Phys.}}   

\allowdisplaybreaks[1]

\begin{document}

\title{Charged Gauss-Bonnet black holes with curvature induced scalarization in the extended scalar-tensor theories}

\author{Daniela D. Doneva}
\email{daniela.doneva@uni-tuebingen.de}
\affiliation{Theoretical Astrophysics, Eberhard Karls University of T\"ubingen, T\"ubingen 72076, Germany}
\affiliation{INRNE - Bulgarian Academy of Sciences, 1784  Sofia, Bulgaria}

\author{Stella Kiorpelidi}
\email{skiorpel@central.ntua.gr}
\affiliation{Department of
Physics, National Technical University of Athens, Zografou Campus
GR 157 73, Athens, Greece}

\author{Petya G. Nedkova}
\email{pnedkova@phys.uni-sofia.bg}
\affiliation{Department of Theoretical Physics, Faculty of Physics, Sofia University, Sofia 1164, Bulgaria}

\author{Eleftherios Papantonopoulos}
\email{lpapa@central.ntua.gr}
\affiliation{Department of Physics,
National Technical University of Athens, Zografou Campus GR 157
73, Athens, Greece}

\author{Stoytcho S. Yazadjiev}
\email{yazad@phys.uni-sofia.bg}
\affiliation{Theoretical Astrophysics, Eberhard Karls University of T\"ubingen, T\"ubingen 72076, Germany}
\affiliation{Department of Theoretical Physics, Faculty of Physics, Sofia University, Sofia 1164, Bulgaria}
\affiliation{Institute of Mathematics and Informatics, Bulgarian Academy of Sciences, Acad. G. Bonchev Street 8, Sofia 1113, Bulgaria}


\begin{abstract}
Recently new scalarized black hole solutions were constructed in the extended scalar-tensor-Gauss-Bonnet gravity, where the scalar field is sourced by the curvature of the spacetime via the Gauss-Bonnet invariant. A natural extension of these results is to consider the case of nonzero black hole charge. In addition we have explored a large set of coupling functions between the Gauss-Bonnet invariant and the scalar field, that was not done until now even in the uncharged case, in order to understand better the behavior of the solutions and the deviations from pure general relativity. The results show that in the case of nonzero black hole charge two bifurcation points can exist -- one at larger masses where the scalarized solutions bifurcated from the Reissner-Nordstr\"om one, and one at smaller masses where the scalar charge of the solutions decreases again to zero and the branch merges again with the GR one. All of the constructed scalarized branches do not reach an extremal limit. We have examined the entropy of the black holes with nontrivial scalar field and it turns out, that similar to the uncharged case, the fundamental branch which possesses scalar field with no nodes is thermodynamically favorable over the  Reissner-Nordstr\"om one for the considered coupling functions, while the rest of the branches possessing scalar field with one or more zeros have lower entropy compared to the GR case and they are supposed to be unstable.
\end{abstract}


\maketitle

\section{Introduction}

The investigation of various modified theories of gravity is motivated by purely theoretical reasons, such as overcoming certain intrinsic inconsistencies of General Relativity (GR). On the other hand, to explain the recent observational results  dark matter and dark energy were considered and  to have a viable theory of Gravity  short and large distance modifications of GR have to be introduced \cite{Joyce:2014kja,Barack_2018}. With the detection of gravitational waves, the modern observational era provides  a new channel for testing gravitational theories in the strong-field regime, and differentiating between them \cite{Abbott:2016blz,Abbott:2016nmj,Abbott:2017vtc,Abbott:2017oio,TheLIGOScientific:2017qsa}. Therefore, it is particularly important to gain intuition about the properties of the compact objects predicted by different modified theories, and the observational signatures, which they can introduce.

Some of the  simplest and viable modifications of GR are the scalar-tensor theories \cite{Barack_2018,Berti_2015}. The presence of a scalar field coupled to gravity
results to hairy black hole solutions. Then powerful no-hair theorems were developed constraining the possible black hole solutions within them. One of the first hairy black hole solutions in an asymptotically flat spacetime was discussed in \cite{BBMB} but soon it was realized that the scalar field diverges on the event horizon and, furthermore, the solution is unstable \cite{bronnikov}, so there is no violation of the no-hair theorems.

The easiest way to make the scalar field regular on the horizon and giving hair to the black hole is
to introduce a scale in the gravity sector of the theory through a
cosmological constant. The resulting black hole solutions with the
presence of the cosmological constant have regular scalar field on
the horizon and all the possible infinities are hidden behind the
horizon.

In the case of a positive cosmological constant with a
minimally coupled or non-minimally coupled scalar field with a self-interaction potential
black hole solutions were found \cite{Zloshchastiev:2004ny,Torii:1998ir,martinez}, but they were shown to be
unstable~\cite{phil,dotti}. In the case of a
negative cosmological constant, stable solutions were found
numerically  \cite{Torii:2001pg,
Winstanley:2002jt} and an exact solution in asymptotically AdS
space with hyperbolic geometry was presented in
\cite{Martinez:2004nb} and generalized later to include charge
\cite{Martinez:2005di} while a generalization to
non-conformal solutions was discussed in \cite{Kolyvaris:2009pc}. Since then a plethora of hairy black hole solutions were reported
of a scalar field coupled to gravity with various self-interacting potentials evading the no-hair theorems.

No-hair theorems  can also be evaded by considering black holes interacting with  matter fields \cite{Stefanov_2008}-\cite{Myung2018}.
In such cases black holes can support a non-trivial scalar field in their exterior region. A similar process is observed also for neutron stars, called spontaneous scalarization \cite{Damour_1993}. For certain ranges of the neutron star densities  a phase transition is realized, so that non-trivial scalar field configurations occur, which are energetically more favorable compared to the corresponding GR solutions.

Hairy black holes can also be solutions of gravity theories in which matter is kinetically coupled to curvature. These theories belong to  general
scalar-tensor Horndeski theories. A gravity model was considered with a scalar
field coupled to the Einstein tensor, and by calculating the quasinormal
spectrum of scalar perturbations, an instability was found outside
the horizon of a Reissner-Nordstr\"om black hole
\cite{Chen:2010qf}. This effect was investigated in details in \cite{Kolyvaris:2011fk}.
A gravity model was considered  consisting of an
electromagnetic field and a scalar field coupled to the Einstein
tensor with vanishing cosmological constant. It was shown that the
Reissner-Nordstr\"om black hole undergoes a second-order phase
transition  to a hairy black hole configuration of generally
anisotropic hair at a certain critical temperature.  The properties of the  hairy black hole
configuration near the critical temperature were calculated perurbatively and it was shown that it is
energetically favorable over the corresponding
Reissner-Nordstr\"om black hole. This ``Einstein hair" is the result of
evading the no-hair theorem thanks to the presence of the derivative
coupling of the scalar field to the Einstein tensor. Furthermore the properties of this phase transition outside the horizon of the
Reissner-Nordstr\"om black hole in asymptotically flat spacetimes were investigated in   \cite{Kolyvaris:2013zfa}.

Another way to obtain  black hole solutions with nontrivial scalar hair and without considering any matter sources is to couple the scalar field directly  to second order algebraic curvature invariants.  In this case the scalar hair is maintained by the interaction with the spacetime curvature. In particular, the extended scalar-tensor-Gauss-Bonnet gravity (ESTGB) was studied in this respect, for which the scalar field is coupled to the Gauss-Bonnet invariant in four dimensions. Compact objects in different classes of ESTBG theories were studied extensively in the literature \cite{Mignemi_1993}-\cite{Kleihaus_2016a}.

Recently within the ESTGB gravity theories it was shown that for certain classes of the coupling function we have spontaneous scalarization of black holes evading in this way  the no-hair theorems \cite{Doneva_2018a}-\cite{Antoniou_2018a}. It was found  that below a certain critical mass the Schwarzschild black hole becomes unstable as a solution in the ESTBG theory \cite{Doneva_2018a,Silva_2018,Myung_2018b}, and new branches of scalarized black holes develop at certain masses. The scalarized branches form a discrete family of solutions labeled by the number of nodes, or zeroes of the scalar field. For each particular branch, however, the scalar charge is not an independent parameter, but it is determined by the black hole mass. Hence, these black holes are characterized by a secondary hair. Investigation of the linear stability with respect to radial perturbations showed that the fundamental branch is stable for certain choices of the coupling function except for very small masses, while the higher order branches are always unstable \cite{Blazquez}.   Such spontaneous scalarization in ESTGB gravity   for neutron stars was also observed in \cite{Silva_2018}, \cite{Doneva_2018}.

In this work we study the scalarization of black holes in ESTGB gravity theory when an electromagnetic field is present.  We construct numerically charged black hole solutions for several scalar field coupling functions. Scalarization is controlled by the spacetime curvature, and occurs for small black hole masses. Below a certain critical mass an instability of the  Reissner-Nordstr\"{o}m solution sets in, and scalarized branches bifurcate from it. The scalarized solutions are thermodynamically more stable than their GR counterparts since they possess a larger entropy. Increasing the black hole charge, the bifurcation point of the scalarized solutions shifts to larger masses.  At the same time, however, the scalarized branches get shorter and the domain of existence of the scalarized solutions in the parametric space decreases. An interesting results is that in the case of nonzero charge a second bifurcation point can exist at low masses, since the scalarized branches can merge again with the Reissner-Nordstr\"{o}m solution. In general, the deviation of the scalarized black holes from GR decreases for larger charges, which is demonstrated by smaller differences in their horizon area and the entropy, and lower maximal value of the scalar charge allowed on the scalarized branches.

The paper is organized as follows. In section II we describe the particular class of ESTGB theories, which we investigate, and study the linear stability of the Reissner-Nordstr\"{o}m black hole as a solution in the theories under consideration. In section III we present the obtained scalarized black holes and analyze their properties. The paper ends with Conclusions.

\section{Gauss-Bonnet theory coupled to a scalar field in the presence of an electromagnetic field}

We consider ESTGB theories coupled to an electromagnetic field with action
\begin{eqnarray}
S=\frac{1}{16\pi}\int d^4x \sqrt{-g}
\Big[R - 2\nabla_\mu \varphi \nabla^\mu \varphi  + f(\varphi)\lambda^2 {\cal R}^2_{GB} + F_{\mu\nu}F^{\mu\nu} \Big]~,\label{eq:quadratic}
\end{eqnarray}
where $\varphi$ is a neutral scalar field, $F_{\mu\nu}$ is the Maxwell tensor, and ${\cal R}^2_{GB}=R^2 - 4 R_{\mu\nu} R^{\mu\nu} + R_{\mu\nu\alpha\beta}R^{\mu\nu\alpha\beta}$ is the Gauss-Bonnet invariant. We assume that the scalar field coupling function $f(\varphi)$ depends only on $\varphi$, while $\lambda$ is the Gauss-Bonnet coupling constant, which has a dimension of length. The action leads to the following field equations
\begin{eqnarray}\label{FE}
&&R_{\mu\nu}- \frac{1}{2}R g_{\mu\nu} + \Gamma_{\mu\nu}= 2\nabla_\mu\varphi\nabla_\nu\varphi -  g_{\mu\nu} \nabla_\alpha\varphi \nabla^\alpha\varphi + 2\left(F_{\mu\alpha}F_{\nu}^{\alpha} -\frac{1}{4}g_{\mu\nu}F_{\alpha\beta}F^{\alpha\beta}\right)~, \nonumber\\
&&\nabla_\alpha\nabla^\alpha\varphi= - \frac{\lambda^2}{4} \frac{df(\varphi)}{d\varphi} {\cal R}^2_{GB}~,\\
&&\nabla_{\mu}F^{\mu\nu}=0~, \nonumber \\
&&\nabla_{[\mu} F_{\alpha\beta]}=0~,  \nonumber
\end{eqnarray}
where  $\nabla_{\mu}$ is the covariant derivative with respect to the spacetime metric $g_{\mu\nu}$ and  $\Gamma_{\mu\nu}$ is defined by
\begin{eqnarray}
\Gamma_{\mu\nu}&=& - R(\nabla_\mu\Psi_{\nu} + \nabla_\nu\Psi_{\mu} ) - 4\nabla^\alpha\Psi_{\alpha}\left(R_{\mu\nu} - \frac{1}{2}R g_{\mu\nu}\right) +
4R_{\mu\alpha}\nabla^\alpha\Psi_{\nu} + 4R_{\nu\alpha}\nabla^\alpha\Psi_{\mu} \nonumber \\
&& - 4 g_{\mu\nu} R^{\alpha\beta}\nabla_\alpha\Psi_{\beta}
 + \,  4 R^{\beta}_{\;\mu\alpha\nu}\nabla^\alpha\Psi_{\beta}
\end{eqnarray}
with
\begin{eqnarray}
\Psi_{\mu}= \lambda^2 \frac{df(\varphi)}{d\varphi}\nabla_\mu\varphi~.
\end{eqnarray}
We are interested in static and spherically symmetric black hole solutions with  scalar and eletromagnetic fields possessing the same symmetries. Taking into account the spacetime symmetries we can use the following ansatz for the metric
\begin{eqnarray}
ds^2= - e^{2\Phi(r)}dt^2 + e^{2\Lambda(r)} dr^2 + r^2 (d\theta^2 + \sin^2\theta d\phi^2 )~,
\end{eqnarray}
while the Maxwell 2-form is
\begin{eqnarray}\label{EM}
F = -\frac{Q}{r^2}dt\wedge dr~,
\end{eqnarray}
where $Q$ is the electric charge. Then, the field equations (\ref{FE}) reduce to the form

\begin{eqnarray}
&&\frac{2}{r}\left[1 +  \frac{2}{r} (1-3e^{-2\Lambda})  \Psi_{r}  \right]  \frac{d\Lambda}{dr} + \frac{(e^{2\Lambda}-1)}{r^2}
- \frac{4}{r^2}(1-e^{-2\Lambda}) \frac{d\Psi_{r}}{dr} - \left( \frac{d\varphi}{dr}\right)^2  - e^{-2\Phi}\frac{ Q^2 }{r^4} =0~, \label{DRFE1} \nonumber \\ && \nonumber \\
&&\frac{2}{r}\left[1 +  \frac{2}{r} (1-3e^{-2\Lambda})  \Psi_{r}  \right]  \frac{d\Phi}{dr} - \frac{(e^{2\Lambda}-1)}{r^2} - \left(
\frac{d\varphi}{dr}\right)^2  + e^{-2\Phi}\frac{{\cal Q}^2 }{r^4}=0~,\label{DRFE2}\\ && \nonumber \\
&& \frac{d^2\Phi}{dr^2} + \left(\frac{d\Phi}{dr} + \frac{1}{r}\right)\left(\frac{d\Phi}{dr} - \frac{d\Lambda}{dr}\right)  +
\frac{4e^{-2\Lambda}}{r}\left[3\frac{d\Phi}{dr}\frac{d\Lambda}{dr} - \frac{d^2\Phi}{dr^2} - \left(\frac{d\Phi}{dr}\right)^2 \right]\Psi_{r}
\nonumber \\
&& \hspace{0.5cm} - \frac{4e^{-2\Lambda}}{r}\frac{d\Phi}{dr} \frac{d\Psi_r}{dr} + \left(\frac{d\varphi}{dr}\right)^2 - e^{-2\Phi}\frac{{\cal Q}^2 }{r^4}=0~, \label{DRFE3}\\ && \nonumber \\
&& \frac{d^2\varphi}{dr^2}  + \left(\frac{d\Phi}{dr} \nonumber - \frac{d\Lambda}{dr} + \frac{2}{r}\right)\frac{d\varphi}{dr} \nonumber \\
&& \hspace{0.5cm} - \frac{2\lambda^2}{r^2} \frac{df(\varphi)}{d\phi}\Big\{(1-e^{-2\Lambda})\left[\frac{d^2\Phi}{dr^2} + \frac{d\Phi}{dr} \left(\frac{d\Phi}{dr} -
\frac{d\Lambda}{dr}\right)\right]    + 2e^{-2\Lambda}\frac{d\Phi}{dr} \frac{d\Lambda}{dr}\Big\} =0~, \label{DRFE4} \\ \nonumber
\end{eqnarray}
where
\begin{eqnarray}
\Psi_{r}=\lambda^2 \frac{df(\varphi)}{d\varphi} \frac{d\varphi}{dr}~.
\end{eqnarray}

The ESTGB theories can possess different properties depending on the form of the scalar field coupling function $f(\varphi)$. In order for the theory to admit the Reissner-Nordstr\"{o}m black hole as a background solution, and further admit scalarization of the black hole solutions,  the coupling function should satisfy the  conditions $\frac{df}{d\varphi}(0)=0$ and $b^2=\frac{d^2f}{d\varphi^2}(0)>0$, where $b$ is a constant, and we assume that the scalar field vanishes at infinity.  The constant $b$ can be always normalized to unity, as this corresponds to the freedom to rescale the Gauss-Bonnet parameter $\lambda$. We can also impose the condition $f(0)=0$, since the field equations are invariant under the shift $f(\varphi)\to f(\varphi) + const$.

\begin{figure}[htb]
\includegraphics[scale=0.5]{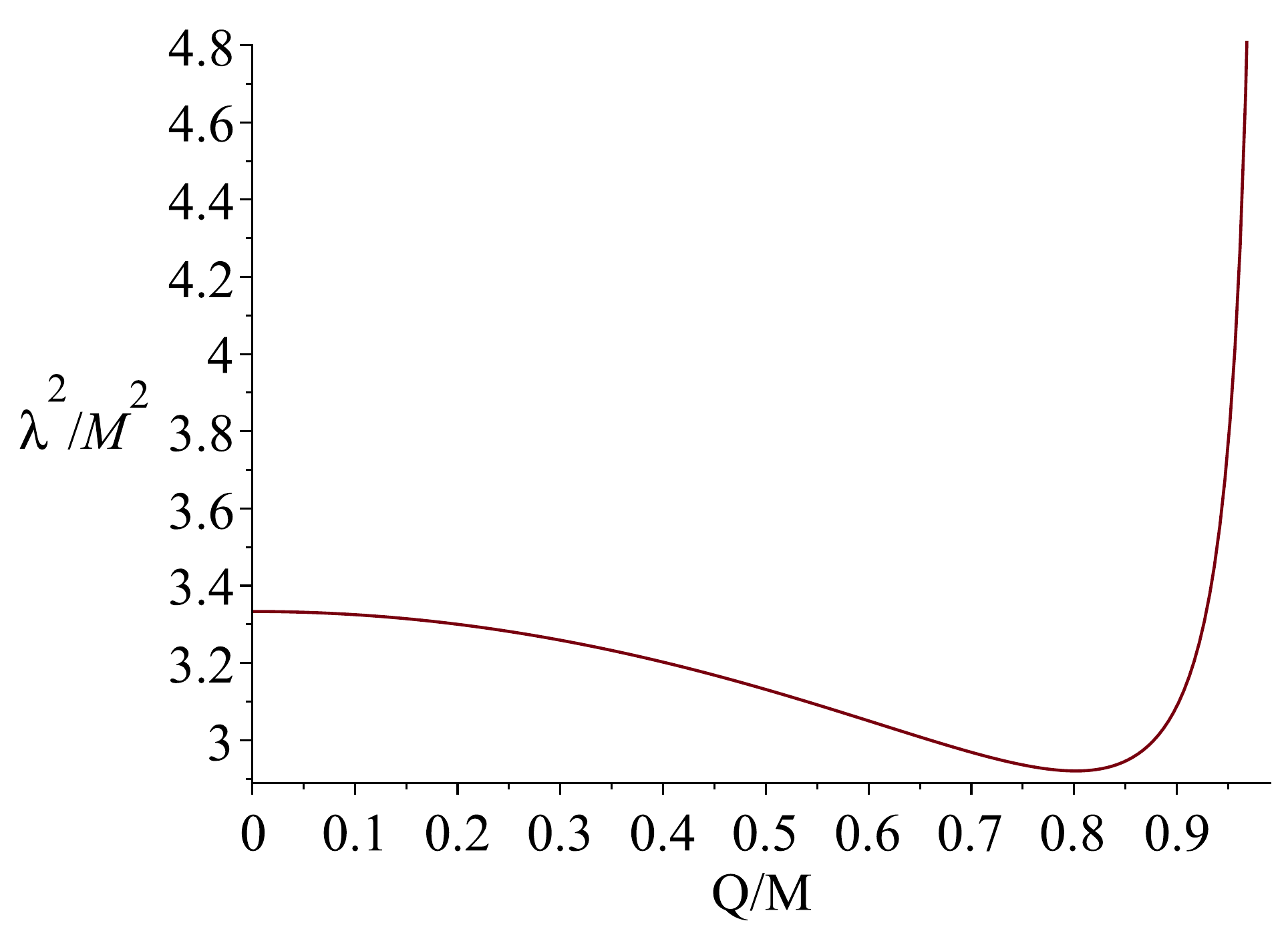}\hspace{0.5cm}
		\caption{Behavior of lower limit of the dimensionless Gauss-Bonnet coupling constant $\tilde\lambda^2$ sufficient to ensure instability of the Reissner-Nordstr\"{o}m  black hole. }
	\label{fig:lambda}
\end{figure}

If the scalar field is vanishing $\varphi=0$, then the field equations admit the  Reissner-Nordstr\"{o}m black hole as a solution.  Scalarized charged black holes are expected to appear in the regions of the parametric space where this solution becomes unstable, forming different branches, which bifurcate from it. Therefore, we will examine its linear stability within the ESTGBT theory under consideration. We consider perturbations of the metric and the scalar field for the Reissner-Nordstr\"{o}m solution with mass $M$ and charge $Q$
\begin{eqnarray}\label{RN}
ds^2 &=& -f(r)dt^2 + \frac{dr^2}{f(r)} + r^2\left( d\theta^2 + \sin^2\theta d\phi^2\right)~, \nonumber \\
f(r) &=& 1 -\frac{2M}{r} + \frac{Q^2}{r^2}~.
\end{eqnarray}
The equations for the metric $\delta g_{\mu\nu}$ and electromagnetic perturbations $\delta A_{\mu}$ decouple from the equation for the scalar field perturbations $\delta \varphi$, and coincide with those in the Einstein-Maxwell gravity. Therefore, the stability is determined from the equation for the scalar perturbations, which takes the form \cite{Doneva_2018a}
\begin{eqnarray}\label{PESF}
\Box_{(0)} \delta\varphi + \frac{1}{4}\lambda^2  {\cal R}^2_{GB(0)} \delta\varphi=0~,
\end{eqnarray}
where $\Box_{(0)}$ and ${\cal R}^2_{GB(0)}$ are the D'alambert operator and the Gauss-Bonnet invariant for the Reissner-Nordstr\"{o}m geometry. In a static and spherically symmetric spacetime we can separate the variables in the standard way $\delta\varphi= \frac{u(r)}{r} e^{-i\omega t}Y_{lm}(\theta,\phi)$ by means of the spherical harmonics $Y_{lm}(\theta,\phi)$. We substitute this expression in Eq. ($\ref{PESF}$) and  introduce the tortoise coordinate $dr_{*}=\frac{dr}{f(r)}$, where $f(r)$ is the Reissner-Nordstr\"{o}m  metric function. Then, the perturbation equation for the scalar field can be reduced to the Schr\"odinger-like equation
\begin{eqnarray}
\frac{d^2u}{dr^2_{*}} + [ \omega^2 - U(r)]u=0 \label{eq:PerturbEq}~,
 \end{eqnarray}
 with an effective potential
 \begin{eqnarray}\label{eq:potential}
U(r) = f(r)\left[\frac{2M}{r^3}- \frac{2Q^2}{r^4} + \frac{l(l+1)}{r^2} - 2\lambda^2\left(\frac{5Q^4}{r^8}-\frac{12MQ^2}{r^7} + \frac{6M^2}{r^6}\right)\right]~.
\end{eqnarray}
A sufficient condition for the existence of an unstable mode \cite{Buell_1995} is that outside of the Reissner-Nordstr\"{o}m black hole horizon the potential should develop a negative well
\begin{equation}
\int_{-\infty}^{+\infty} U(r_*)dr_* = \int_{r_H}^\infty \frac{U(r)}{f(r)} dr <0~,
\end{equation}
where $r_H = M + \sqrt{M^2-Q^2}$ is the horizon radius of the Reissner-Nordstr\"{o}m black hole. Spherical symmetry requires that we consider only the zero mode $l=0$. Normalizing the charge and the Gauss-Bonnet coupling constant to the black hole mass $\tilde Q = Q/M$ and $\tilde\lambda = \lambda/M$, the sufficient condition leads to the inequality \footnote{The normalization used in the equations below is done for convenience and is different from the one employed in the results section, where the quantities are normalized to the parameter $\lambda$ instead. }
\begin{eqnarray}
\left(1+2\sqrt{1-\tilde Q^2}\right)\left(1+\sqrt{1-\tilde Q^2}\right)^3 - \frac{6}{35}\tilde \lambda^2\left(\left(2+5\sqrt{1-\tilde Q^2}\right)^2 - 7\right)<0~.
\end{eqnarray}
A necessary condition for this inequality to be satisfied is
\begin{equation}\label{RN_st}
\left(2+5\sqrt{1-\tilde Q^2}\right)^2 - 7 >0~,
\end{equation}
which constrains the charge to mass ratio in the range $|Q|/M < \frac{1}{5}\sqrt{14 + 4\sqrt{7}}\approx 0.9916$. Provided that the inequality ($\ref{RN_st}$) is satisfied, the Reissner-Nordstr\"{o}m solution becomes unstable if its charge to mass ratio further obeys the relation
\begin{eqnarray}
 f(\tilde Q) = \frac{35}{6}\frac{\left(1+2\sqrt{1-\tilde Q^2}\right)\left(1+\sqrt{1-\tilde Q^2}\right)^3}{\left(2+5\sqrt{1-\tilde Q^2}\right)^2 - 7} < \tilde\lambda^2~.
\end{eqnarray}
The qualitative behavior of the function in the left-hand side of this relation is presented in Fig. $\ref{fig:lambda}$ showing the lower limit of the dimensionless Gauss-Bonnet parameter $\tilde\lambda =\lambda/M$, which is sufficient to ensure instability for a particular charge to mass ratio.

The Gauss-Bonnet coupling $\lambda$ shows how strong is the coupling of the scalar field to the spacetime curvature. Then Fig. $\ref{fig:lambda}$ shows
that a strong attractive gravitational force acting on the scalar field destabilizes the  Reissner-Nordsr\"{o}m black hole. An interesting  case is if the scalar field is charged. Then there are two competing forces -- the gravitational attraction and the electromagnetic repulsion. Then the background Reissner-Nordsr\"{o}m black hole will be destabilized and scalar hair will be formed outside its horizon as the result of the balance of these two forces. 

To see such an effect we will study how a charged massive scalar field coupled to the GB term is superradiantly amplified. We will show that the field scattered off the horizon of the background  Reissner-Nordstr\"om black hole will be trapped in a potential well and it will be superradiant   amplified leading to an instability of the background Reissner-Nordstr\"om black hole. Consider a charged massive scalar field coupled to the GB term as in the action (\ref{eq:quadratic}). Then the Klein-Gordon equation reads

\begin{equation}
\left[ \left( \nabla^\nu - i q A^\nu \right) \left( \nabla_\nu - i q A_\nu \right) - \mu^2 + \frac{1}{4} \lambda^2 \mathcal{R}^2_{GB} \right] \phi =0~,
\end{equation}
where $A_\nu$ is the gauge potential and in our case the only nonzero component is $A_t=-Q/r$, while $\mu$ and $q$ are the mass and the charge of the scalar field.
After the decomposition of the scalar field as
\begin{equation}
\phi_{lm} (t,r,\theta , \phi) = u(r) e^{-i \omega t} Y_{lm}(\theta, \phi )~,
\end{equation}
 the radial part of the Klein-Gordon equation is given by
\begin{equation}
\Delta \frac{d}{dr} \left( \Delta \frac{du}{dr} \right) + U u = 0~,
\end{equation}
where
\begin{equation}
\Delta = r^4 f(r)~,
\end{equation}
and the potential is given by
\begin{equation}
U = r^4 \left(\omega r^2 -qQr \right)^2 - \Delta r^2 \left[ \mu^2 r^2+l(l+1)-\frac{1}{4} r^2 \lambda^2 \mathcal{R}^2_{GB} \right]~.
\end{equation}
We are interested in solutions of the radial equation with the physical boundary conditions of purely ingoing waves at the black hole horizon and a decaying bounded solution at spatial infinity. So by doing the transformation
\begin{equation}
\frac{\partial }{\partial r}= \frac{1}{f(r)} \frac{\partial }{\partial r^*}~,
\end{equation}
where $r^*$ is the tortoise radial coordinate we get the solutions at the horizon and at spatial infinity of a Schrodinger-like equation as
\begin{align}
u(r^* \to - \infty ) &\sim e^{r_H^2 \left(\omega r^2 -qQr \right) r^*}~,\label{exten} \\
u(r^* \to +\infty ) & \sim e^{- \sqrt{\mu^2-\omega^2} r^*}~,
\end{align}
where $r_H$ is the horizon of the Reissner-Nordstr\"om black hole. As it can be seen from (\ref{exten}) if
\begin{equation}
\omega < \frac{qQ}{r_H}~,\label{supcon}
\end{equation}
then there is extraction of energy from the Reissner-Nordstr\"om black hole. This is known as the superradiance condition. However, this condition is independent on the GB coupling $\lambda$ and therefore it does not have the information that the scalar wave is coupled to spacetime curvature. To have
an instability of the background Reissner-Nordstr\"om black hole this energy has to be trapped in a potential well outside its horizon \cite{Hod:2013nn,Hod:2015hza,Hod:2016kpm,Kolyvaris:2017efz,Kolyvaris:2018zxl}. Note that the well itself must be separated from the black hole horizon by a potential barrier. As in the case of the neutral scalar field this potential well is provided by the coupling of the scalar field to the GB term.

It is convenient to define a new radial function $v$  by
\begin{equation}
v=\sqrt{\Delta} u~,
\end{equation}
in terms of which the radial equation can be written in the form of a Schrodinger-like wave equation
\begin{equation}
\frac{d^2 v}{dr^2} + (\omega^2-V ) v=0~,
\end{equation}
where
\begin{equation}
\omega^2-V=\frac{1}{\Delta^2} \left(U+\frac{1}{4} \frac{d \Delta}{dr} - \frac{1}{2} \frac{d^2 \Delta}{dr^2} \right)~.
\end{equation}
So we have to analyze the behaviour of this effective potential $V(r;M,Q,\mu , q ,\omega,l,\lambda)$. The effective potential is characterized by a set of seven parameters : $\left\lbrace M, Q, \mu ,q, \omega ,l,\lambda \right\rbrace $.

\begin{figure}[h!]
\centering
\includegraphics[scale=0.6]{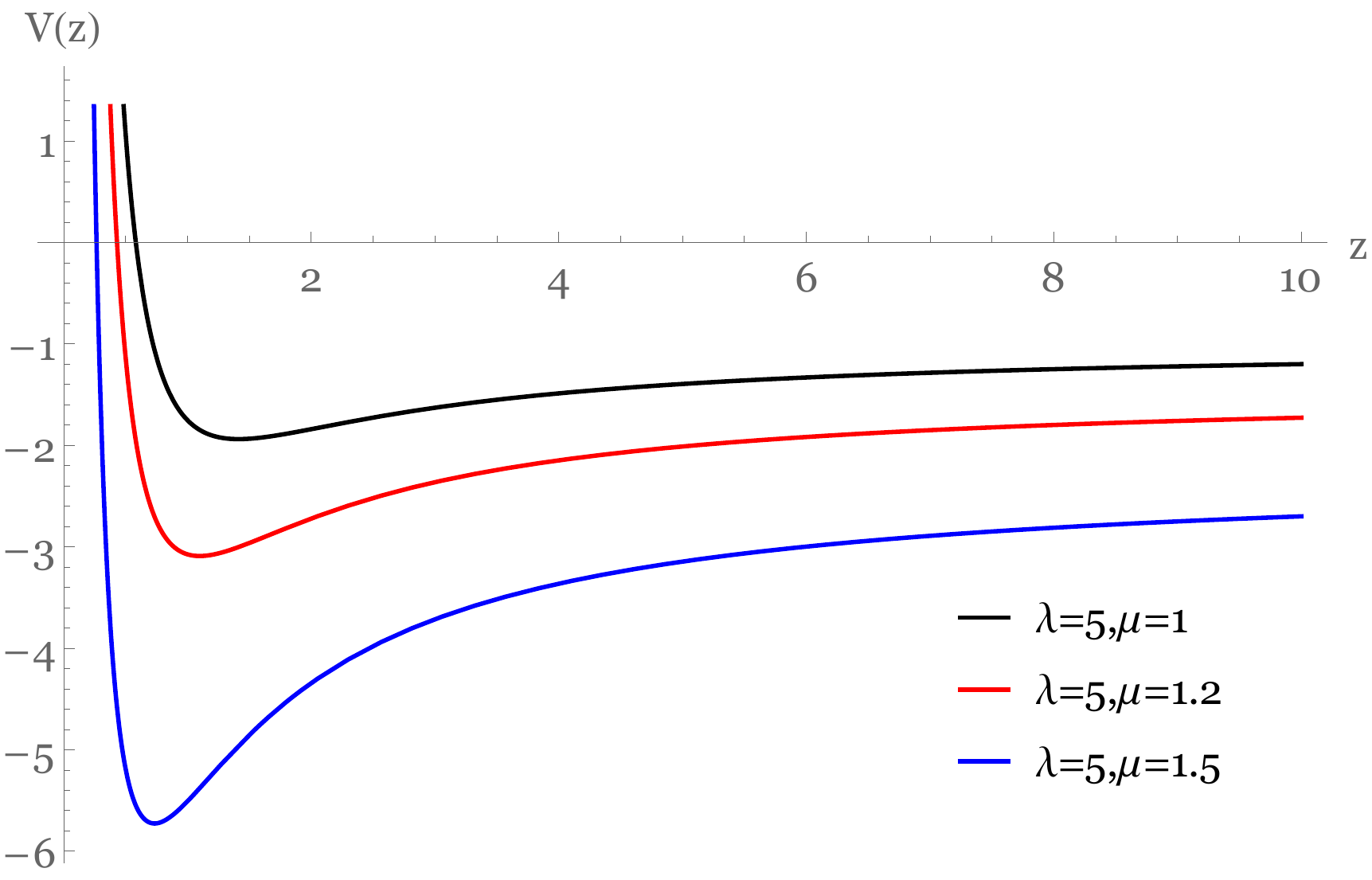}
\caption{The effective potential as a function of the coordinate $z$ for $\lambda=5$ and for various values of the mass $\mu$. In the superradiant regime we set $M=1$, $Q=\dfrac{1}{2}$, $q=\dfrac{1}{2}$, $\omega=0.01$ and $r_H=M+\sqrt{M^2-Q^2}$. } \label{plot1}
\end{figure}

\begin{figure}[h!]
\centering
\includegraphics[scale=0.6]{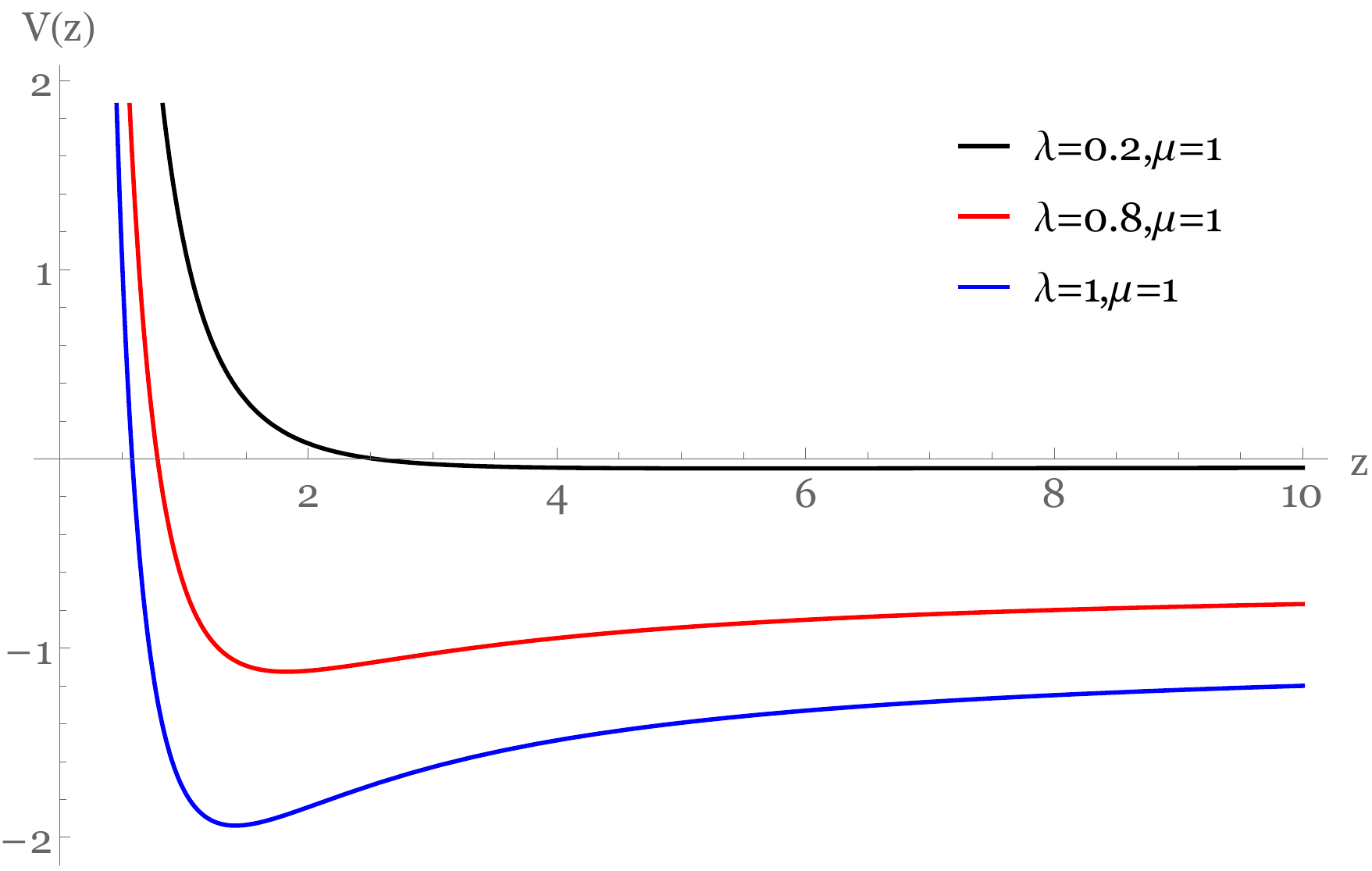}
\caption{The effective potential as a function of the coordinate $z$ for $\mu=1$ and for various values of the coupling $\lambda$. In the superradiant regime we set $M=1$, $Q=\dfrac{1}{2}$, $q=\dfrac{1}{2}$, $\omega=0.01$ and $r_H=M+\sqrt{M^2-Q^2}$.} \label{plot2}
\end{figure}

Firstly we can check that there is a potential barrier at the black hole horizon
\begin{equation}
V(r \to r_H )=\infty~.
\end{equation}
The analysis of the effective potential is a really hard task because its form is very complicated.
It is convenient to define a new variable
\begin{equation}
z=r-r_H~.
\end{equation}
Then if we plot the effective potential we can see that in the regime of the superradiant condition (\ref{supcon}) (specific values for parameters $Q$, $q$, $\omega$) this effective potential can take the form of a potential well outside the black hole as it is shown in Fig.~\ref{plot1} and Fig.~\ref{plot2}.
Observe that as the GB coupling is increased the potential well is increased and this will lead to an instability of the background Reissner-Nordstr\"om black hole.

The performed stability analysis in both cases of a neutral or charged scalar field provides only a sufficient condition for instability and does not exclude scalarization occurring at smaller values of the GB coupling. In the next sections we obtain numerically the scalarized black hole solutions  expected in the region of instability of the background Reisnner-Nordstr\"{o}m  black hole solution
and examine their behavior. We carried out the analysis for the neutral scalar field and we leave the treatment of the charged case for future work.

\section{Scalarization of the Reissner-Nordstr\"{o}m  black hole }

We construct charged black hole solutions by integrating numerically the system of reduced field equations \eqref{DRFE1}--\eqref{DRFE4} using a shooting method. The boundary conditions, which are imposed, are determined by the requirements that the solutions are asymptotically flat and regular on the black hole horizon  $r= r_H$. Asymptotic flatness implies the following asymptotic behavior of the metric functions and the scalar field at spacetime infinity $r\rightarrow \infty$: $\Phi|_{r\rightarrow\infty} \rightarrow 0$, $\Lambda|_{r\rightarrow\infty} \rightarrow 0$,  $\varphi|_{r\rightarrow\infty} \rightarrow 0$. On the other hand, in the near-horizon region the metric functions behave as $e^{2\Phi}|_{r\rightarrow r_H} \rightarrow 0$, $e^{-2\Lambda}|_{r\rightarrow r_H} \rightarrow 0$. We require that the scalar field is regular on the black hole horizon together with its first and second derivatives. This leads to the condition
\begin{align}
\left(\frac{d\varphi}{dr}\right)_{H} &= -\frac{1}{4\,\Phi_1\,r^2_H\,f_1(\varphi)\left(r^6_H - 4Q^2\,f^2_1(\varphi)\,\right)}\bigg[ \left(Q^2 + \Phi_1r^3_H\right)\left(r^6_H - 8Q^2\,f^2_1(\varphi)\,\right) \nonumber \\[2mm]
 &\pm r^2_H\sqrt{r^8_H\left(Q^2 + \Phi_1r^3_H\right)^2 - 8\Phi_1\,r_H\,f^2_1(\varphi)\left(2Q^2 + 3\Phi_1r^3_H\right)\left(r^6_H - 4Q^2\,f^2_1(\varphi)\,\right)}\,\bigg]~,
\end{align}
where we use the notations $\Phi_1 = \left(de^{2\Phi}/dr\right)_{r=r_H}$ and $f_1 = \left(df(\varphi)/d\varphi\right)_{r=r_H}$. We consider only the plus sign, since it recovers the Reissner-Nordsr\"{o}m black hole in the absence of the scalar field. From this expression we obtain that non-trivial scalar field configurations are possible only if the following constraint is satisfied
\begin{eqnarray}\label{eq:ConditionExistence}
f^2_1(\varphi)\left[r^6_H - 4Q^2\,f^2_1(\varphi)\right] < \frac{r^7_H\left(Q^2 + \Phi_1r^3_H\right)^2}{8\Phi_1\left(2Q^2 + 3\Phi_1r^3_H\right)}~.
\end{eqnarray}
The obtained black hole solutions are characterized by three parameters associated with their mass $M$, electric charge $Q$, and scalar charge $D$. They are determined by the Maxwell 2-form ($\ref{EM}$), and the asymptotic expansions of the metric functions and the scalar field at infinity\footnote{Here we assume a zero cosmological value of the scalar field, i.e. $\varphi_\infty=0$.}
\begin{equation}
\Phi|_{r\rightarrow \infty}\approx  -\frac{M}{r} + O(1/r^2),\;\;\; \varphi|_{r\rightarrow \infty}\approx \frac{D}{r} + O(1/r^2)
\end{equation}

We investigate the scalarization of the Reissner-Nordstr\"{o}m  black hole in the ESTGB theory by considering three different scalar field coupling functions. The first coupling function is given by
\begin{eqnarray}
f(\varphi)=  \frac{1}{2\beta} \left[1- \exp(-\beta\varphi^2)\right] \label{eq:caseI}~,
\end{eqnarray}
where  $\beta$  is a constant. In the following analysis it is denoted as \textit{Case I}. The same coupling function with $\beta=6$ was used in the construction of the scalarized Schwarzschild black holes in the ESTGB gravity in \cite{Doneva_2018a} and it was proven in \cite{Blazquez} that the first fundamental branch is stable (except for very small masses) while the rest of the scalarized branches are unstable. The motivation for this particular choice \eqref{eq:caseI} comes also from the fact that the scalarized neutron stars in the standard scalar-tensor theories were mainly studied for such coupling function \cite{Damour_1993}.

Further two coupling functions, which lead to spontaneous scalarization, are introduced, in order to study the sensitivity of the process on the particular theory. The function
\begin{equation}
f(\varphi)=  \frac{1}{\beta^2 } \left[1- \frac{1}{\cosh(\beta\varphi)}\right] \label{eq:caseII}
\end{equation}
is denoted in the results  as \textit{Case II}, while we refer to the function
\begin{eqnarray}
f(\varphi)=  \frac{\varphi^2}{2(1+ \beta^2\varphi^2)} \label{eq:caseIII}
\end{eqnarray}
as \textit{Case III}. As a matter of fact, the coupling function \eqref{eq:caseIII} in the limit $\beta=0$ is the same (up to a scaling factor) as the one used in \cite{Silva_2018}.

All of the three coupling  functions have the same expansion up to a leading order when $\varphi \rightarrow 0$ thus the scalarized solutions would have the same behavior very close to the bifurcation points and the bifurcation points are located at the values of the mass for fixed $Q$. As we will see below, in all of the considered cases decreasing the coupling constant $\beta$ leads to larger deviations from GR but on the other hand if $\beta$ is small enough, the scalarized branches of solutions might become very short or even completely unstable.

As expected by the analysis of the linear stability, the Reissner-Nordstr\"{o}m black hole becomes unstable in certain regions of the parameter space. In these regions we obtain numerically scalarized solutions bifurcating from it at particular masses, and spanning to some other nonzero mass lower than the bifurcation point. In some of the cases the black hole scalar charge smoothly decreases to zero at the end of the branch thus merging again with the Reissner-Nordstr\"{o}m black hole at a second bifurcation points located at a lower mass or the sequences of solutions is terminated before $D=0$ is reached due to violation of condition \eqref{eq:ConditionExistence}.

In general, the scalarized uncharged solutions form a countable family of branches, which are characterized by the number of nodes of the scalar field.   However, all the branches except for the first one, which possesses scalar field with no zeroes, are expected to be unstable, as demonstrated in previous studies \cite{Doneva_2018a}, \cite{Blazquez}. The picture in the charged case is quite similar with an important difference. The results in \cite{Doneva_2018a}, \cite{Blazquez} for $Q=0$ show that the bifurcation points and the additional scalarized branches appear at smaller and smaller masses with the increase of the number of nodes of the scalar field. But for a fixed $Q$ the  Reissner-Nordstr\"{o}m solution possesses a minimum mass determined by the extremal limit. Thus it is natural to expect that the scalarized solutions also have similar lower mass limit that quickly increases with the increase of $Q$ which results in a lower number in bifurcation points. As a matter of fact for most of the cases presented in the paper only the  fundamental scalarized branch exists.

 The stability of the branches with nontrivial scalar field is not supposed to change, though, and all the nontrivial solutions characterized with scalar field that has one or more zeros are expected to be unstable. Therefore, in our analysis we will investigate only the fundamental branch. The location of the bifurcation points depends only on the value of $\lambda$ and not on the particular form of the coupling function, as one can see from the effective potential \eqref{eq:potential}. In addition the three coupling functions we study belong to the same class determined by the conditions $\frac{df}{d\varphi}(0)=0$ and $\frac{d^2f}{d\varphi^2}(0)=1$, which guarantees that they behave identically for small scalar fields and lead to coinciding bifurcation points of the scalarized branches.

\begin{figure}[htb]
\includegraphics[scale=0.7]{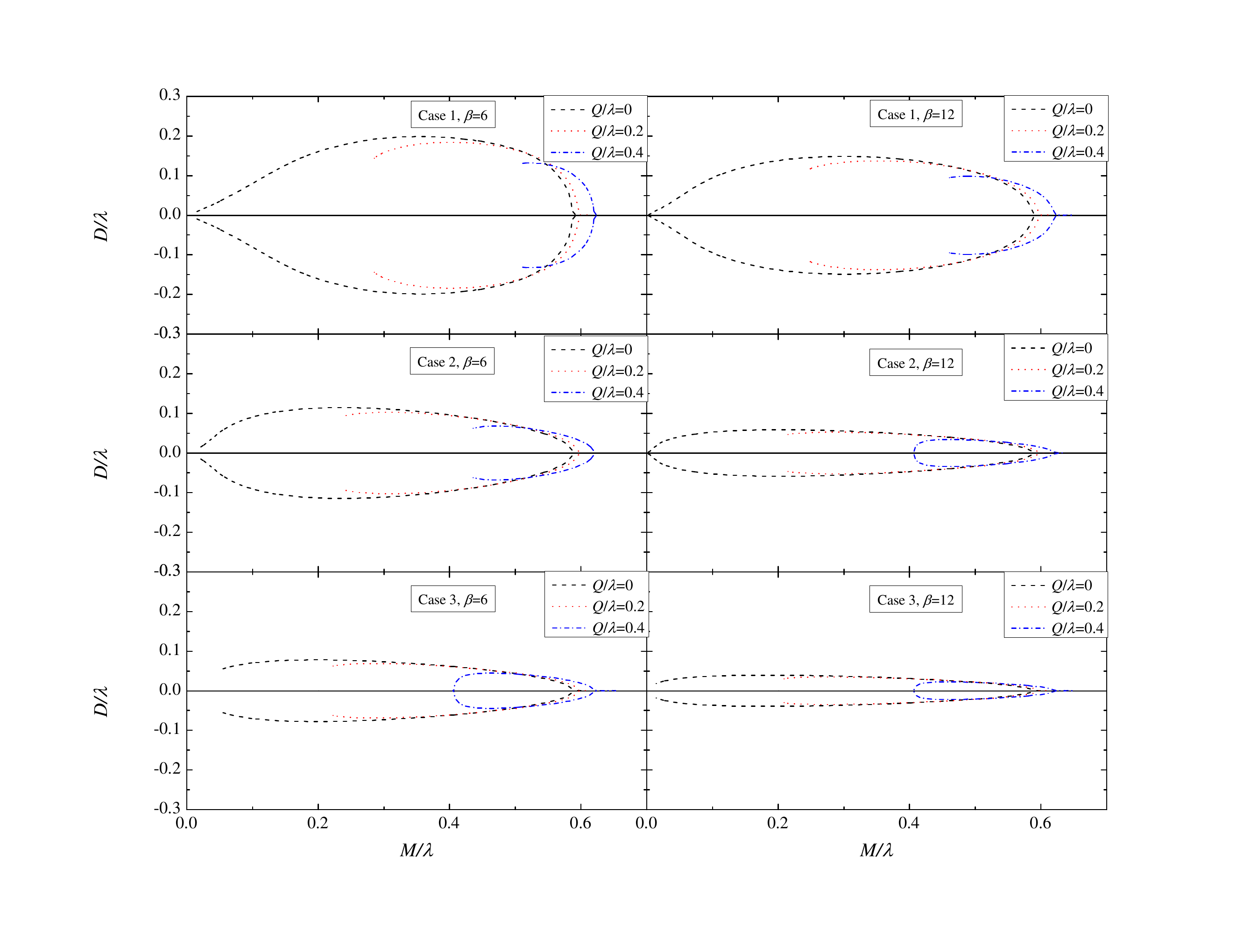}\hspace{0.5cm}
		\caption{The scalar charge of the black hole as a function of its mass. Figures for all three of the coupling functions are shown ({\it Case I}, {\it Case II} and {\it Case III}) for $\beta=6$ and $\beta=12$. In each figure the sequences with black hole charge  $Q/\lambda=0$,  $Q/\lambda=0.2$ and  $Q/\lambda=0.4$ are shown. All the quantities are normalized to the parameter $\lambda$.}
	\label{fig:DM}
\end{figure}

We present the obtained black hole solutions in Fig. $\ref{fig:DM}$, where we show the behavior of the scalar charge as a function of the black hole mass for several values of the electric charge. The results presented in this figure and the one below are normalized to the parameter $\lambda$ as denoted on the graphs and similar to previous studies \cite{Doneva_2018}, \cite{Blazquez}. For each coupling function we study two regimes: strong coupling with larger deviations from GR represented by the value of the coupling constant $\beta = 6$, and weak coupling for $\beta = 12$.  We see that the scalarized solutions are characterized by small deviations from the scalar-free Reissner-Nordstr\"{o}m black hole near the two bifurcation points -- the one at higher mass at the beginning of the branch and the one at lower mass at the end of the branch. The second bifurcation point exists only if condition \eqref{eq:ConditionExistence} is fulfilled for the whole branch, otherwise the branch is terminated at some nonzero scalar charge that happens in {\it Case I} for $\beta=-6$ and $\beta=-12$ and in {\it Case II} for $\beta=-6$ in Fig. $\ref{fig:DM}$. We should note that a second bifurcation point at smaller masses is observed only for nonzero charges as one can see in the figure, while for $Q=0$ the branches are terminated either because of violation of condition \eqref{eq:ConditionExistence} or they reach the $M=0$ limit.

For intermediate masses the scalarized branches can distinguish considerably from the Reissner-Nordstr\"{o}m black hole, the deviation being measured by the value of the scalar charge. Increasing the parameter $\beta$ the maximal deviation for the scalar-free Reissner-Nordstr\"{o}m solution, or equivalently the maximal absolute value of the scalar charge, decreases for all of the coupling functions. Such behavior is consistent with the expected weaker deviation from $GR$ for larger values of the coupling constant $\beta$. It is also interesting to observe that the three coupling functions lead to qualitatively similar results and what differs is only the magnitude of the scalar charge (and thus the deviation from GR) that is practically controlled by the value of $\beta$.

In most cases for a fixed $Q$ the branches become longer (they terminate at larger mass) as $\beta$ increases since condition \eqref{eq:ConditionExistence} is fulfilled for a larger range of masses. It is natural to expect that if $\beta$ is too small, the branches become very short because of violation of condition \eqref{eq:ConditionExistence} shortly after the first bifurcation point at larger masses. For example, in the limit $\beta =0$ and for uncharged black holes ($Q=0$) it was explicitly shown in \cite{Blazquez}  that  in \textit{Case III} even the first fundamental branch of solutions is very short and unstable. The motivation behind using $\beta=6$ as the minimum $\beta$ for our calculations was the fact that even for $Q=0$, $\beta=6$ is more or less close to the limiting value below which one can not obtain ``nice'' branches of solutions with nontrivial scalar field that reach close to the $M=0$ limit.

The influence of the electric charge on the scalarized solutions can be recognized by several effects. Charging up the Schwarzschild black hole leads to shifting the first bifurcation point of the scalarized solutions to larger masses as their values increase with the black hole charge. At the same time the scalarized branches get shorter and they never reach the extremal solution because of violation of condition \eqref{eq:ConditionExistence}.  In result, the branches span in a smaller region in the parametric space, corresponding, however, to higher masses.  Increasing the black hole charge, we observe that the maximum absolute value of the scalar charge for the scalarized branches decreases.

\begin{figure}[htb]
\includegraphics[scale=0.7]{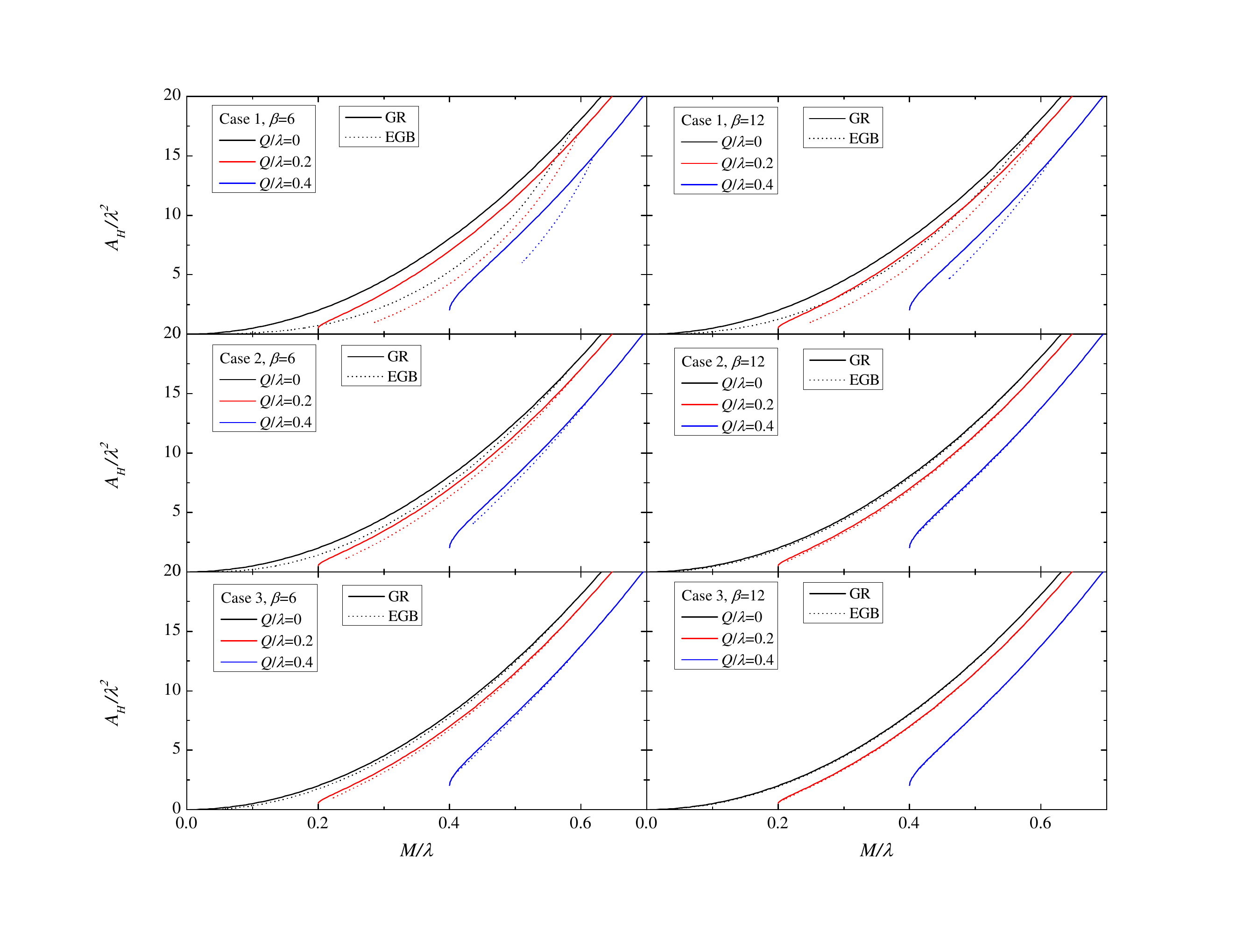}\hspace{0.5cm}
		\caption{The area of the black hole horizon as a function of the mass. The notations are the same as in Fig. \ref{fig:DM}.}
	\label{fig:AH}
\end{figure}

In Fig. $\ref{fig:AH}$  we present the behaviour of the horizon area $A_H=4\pi r_H^2$ as a function of the black hole mass. This quantity can be also viewed as a measurement of the deviation from the  Reissner-Nordstr\"{o}m black hole. Scalarized black holes are smaller than their GR counterparts, as the difference is largest for intermediate masses similar to the scalar charge. As expected, after a comparison with Fig. \ref{fig:DM} one can conclude that the larger the scalar change is, the larger the deviation from GR. In order to have a non-negligible difference, though, the parameter $\beta$ should be small enough.

For the same value of the parameter $\beta$ the different coupling functions lead to a different degree of deviation of the scalarized solutions from GR. \textit{Case I} causes the strongest deviation followed by \textit{Case II} and \textit{Case III} in decreasing order, resulting in the largest absolute value of the scalar charge admitted on the scalarized branches, and the largest discrepancies in the horizon area curves compared to GR. We should note, however, that this is a residual effect of the particular choices of $\beta$ that are used and the particular form of the coupling functions. Decreasing further $\beta$ in \textit{Case II} and \textit{Case III} would produce larger differences with the Reissner-Nordstr\"{o}m black hole that are of similar magnitude to \textit{Case I}.

\begin{figure}[htb]
\includegraphics[scale=0.7]{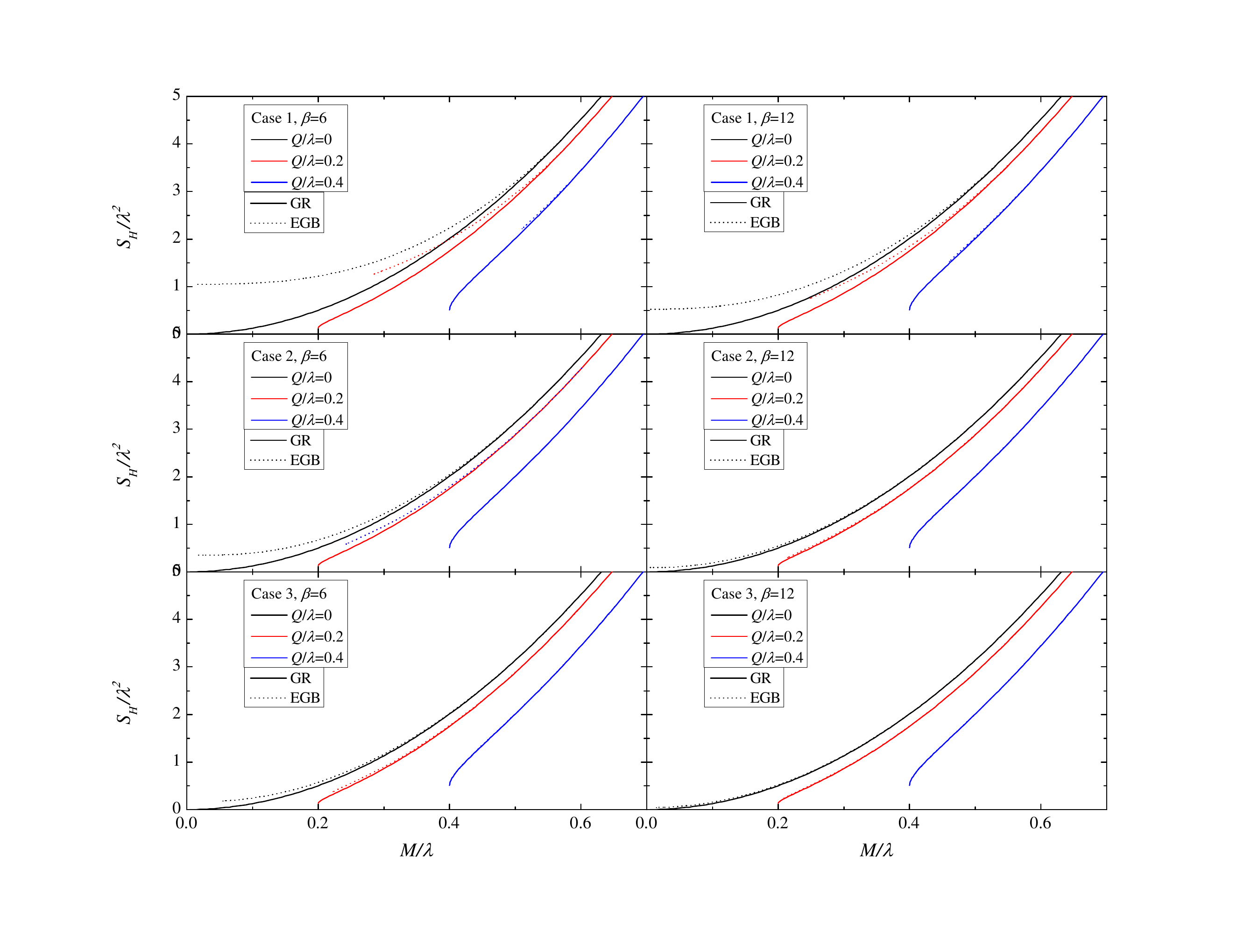}\hspace{0.5cm}
		\caption{Black hole entropy as a  function of its mass. The notations are the same as in Fig. \ref{fig:DM}.}
	\label{fig:SH}
\end{figure}

We further study the entropy of the obtained solutions. When a Gauss-Bonnet invariant is included in the gravitation action, the entropy is not determined solely by the horizon area. We apply the approach of Wald and Iver in \cite{Wald_1993}, \cite{Iyer_1994} valid for any theory of gravity with a diffeomorphism invariant  Langrangian. According to it a Noether charge is associated to the black hole entropy $S_H$, which leads to the expression
\begin{eqnarray}
S_{H}= 2\pi\int_{H} \frac{\partial {\cal L}}{\partial R_{\mu\nu\alpha\beta}} \epsilon_{\mu\nu}\epsilon_{\alpha\beta}~,
\end{eqnarray}
where ${\cal L}$ is the Lagrangian density and  $\epsilon_{\alpha\beta}$ is the volume form on the 2-dimensional cross section $H$ of the horizon.  In our case we obtain the explicit formula $S_H = \frac{1}{4} A_H + 4\pi \lambda^2 f(\varphi_{H})$. The entropy as a function of the black hole mass is illustrated in Fig. $\ref{fig:SH}$. We see that for all the obtained black holes the entropy of the scalarized solutions exceeds that of their GR counterpart. Thus, we can conclude that the scalarized configurations are thermodynamically more favorable and show better stability. However, the deviation between the entropy curves for the scalarized and scalar-free solutions decreases when the black hole charge is increased, and for large charges they approach each other.
We should point out again that the in the figure only the fundamental branches of solutions, possessing scalar field with no nodes, are shown and the conclusions above apply for them. We have checked that similar  to the uncharged $Q=0$ case, the rest of the branches with scalar field possessing one or mode nodes, have lower entropy and they are supposed to be unstable.

\begin{figure}[htb]
\includegraphics[scale=0.7]{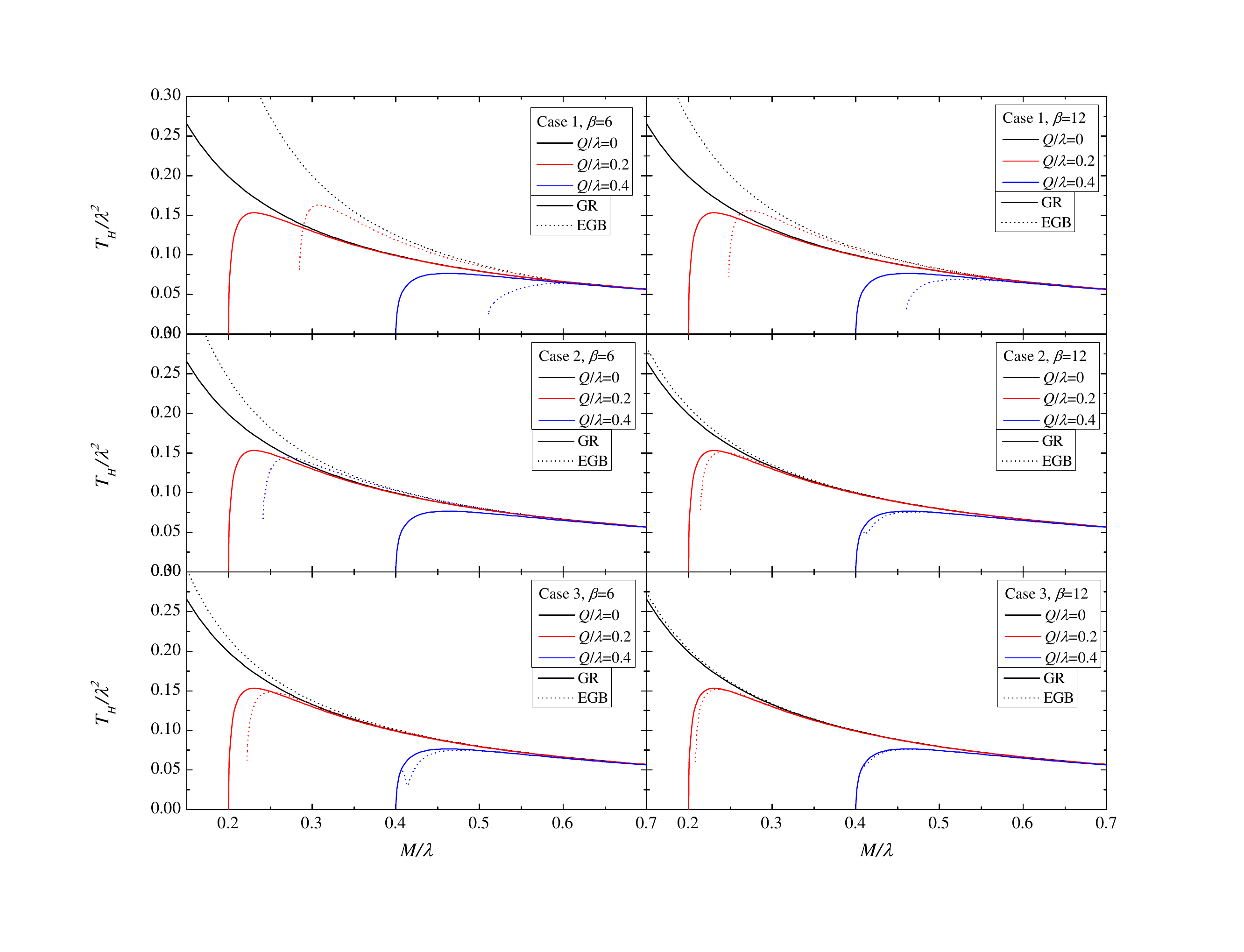}\hspace{0.5cm}
	\caption{The temperature of the black hole as a function of its mass. The notations are the same as in Fig. \ref{fig:DM}.}
	\label{fig:ThM}
\end{figure}

The temperature as a function of the black hole mass is plotted in Fig. \ref{fig:ThM}. As one can see for all of the considered cases the scalarized solutions fail to reach the extremal $T=0$ limit contrary to the Reissner-Nordstr\"{o}m solutions. The reason is either the fact that the branches are terminated because of violation of condition \eqref{eq:ConditionExistence} or because the scalar charge goes to zero at the end of the branch and it mergers again with the Reissner-Nordstr\"{o}m solution for small masses and temperatures. Even though we calculated much larger number of sequences of scalarized black holes, we did not manage to obtain extremal solutions with $T=0$.

\section{Conclusions}

In this work  we have studied scalarized black holes in ESTGB theory when the nontrivial scalar field is sourced by the curvature near the black hole horizon. In particular, we have extended the previous studies on the subject by considering a nonzero charge of the black hole. We have considered only the first branch of black holes possessing a scalar field that has no nodes, since the rest of the branches which have scalar field with one or more nodes, are supposed to be unstable.

Adding a black hole charge leads to some interesting consequences for the scalarized solutions. Naturally, the branches of black holes with nontrivial scalar field do not span up to the $M=0$ limits unlike the $Q=0$ case and at least for all of the studied cases they never reach the extremal solution. Instead the branches are either terminated because of violation of the condition for existence of scalarized solutions \eqref{eq:ConditionExistence}, or they merge with the Reissner-Nordsr\"{o}m solution before the extremal solution is reached. Thus, the Reissner-Nordsr\"{o}m solution can posses two bifurcation points  -- one at larger masses where the scalarized branch emerges from the GR one, and one at smaller masses where the scalarized branch merge again with the GR one. This is different compared to the $Q=0$ case where only the first bifurcation point exists. The numerical results show, that the first bifurcation point is shifted to larger masses as $Q$ increases. As expected, the scalar charge of the scalarized branches tends to zero at the two bifurcation points and maximum deviations for the Reissner-Nordsr\"{o}m black holes is observed for intermediate values of the mass.

In our studies we have employed several coupling functions in order to better determine which effects dependent on the particular choice of the ESTGB theory and to examine the possible deviations from GR. Such a comprehensive study of coupling functions was not done until now even in the $Q=0$ case. All of the coupling functions, though, exhibit the same behavior in the limit of vanishing scalar field, that was chosen intentionally in order for the scalarized black hole solutions to exist. Thus, the bifurcation points are located at  the same values of the mass for a fixed charge. The studies show, that the deviations from the Reissner-Nordsr\"{o}m solution and the conditions for the existence of scalarized solutions \eqref{eq:ConditionExistence} are largely dependent on the parameters of the coupling function and significant differences with GR are possible for appropriately chosen values of the coupling parameters.

We have studied the temperature and the entropy of the black holes as well. Because of the presence of a second bifurcation point for the considered solutions, it is natural to expect that the $T=0$ limit is not reached and extremal solutions are not present. We have investigated numerically a very larger range of parameters none of the constructed branches reach the extremal limit. In addition, the scalarized solutions are thermodynamically more stable compared to the Reissner-Nordsr\"{o}m solution since they have higher entropy. This applies, of course, only to the solutions with nontrivial scalar field which has no zeros, while the solutions with one or more zeros (not presented in the present paper) have lower entropy and therefore they are supposed to be unstable similar to the uncharged $Q=0$ case.

It would be interesting to study the scalarization of the charged Gauss-Bonnet black holes in the extended scalar-tensor theories induced by a charged scalar field. In this case, except for the gravitational attraction there will also be an electromagnetic repulsion. As a first approach to this problem we considered a massive scalar field coupled to the Gauss-Bonnet invariant, scattered off the horizon of the Reissner-Nordsr\"{o}m black hole. We found that a potential well is formed outside the horizon of the black hole trapping the scalar field. This results in the superradiant amplification of the scalar field leading to an instability of the background black hole. We leave for future work the backreaction of the charged scalar field to the metric which will lead to possible scalarized charged Gauss-Bonnet black hole solutions.

\section*{Acknowledgements}
DD would like to thank the European Social Fund, the Ministry of Science, Research and the Arts Baden-W\"{u}rttemberg for the support. DD is indebted to the Baden-W\"{u}rttemberg
Stiftung for the financial support of this research project by the Eliteprogramme for Postdocs. The support by the Bulgarian NSF Grant DFNI T02/6, Sofia University Research
Fund under Grants 80-10-73/2018 and 3258/2017, and
COST Actions CA15117, CA16104, CA16214   is also gratefully acknowledged.


\end{document}